\documentclass[11pt]{article}
\usepackage{amsmath,amssymb,amsthm,amsxtra,overpic,bbm,bm,epsfig,subfigure}
\usepackage{color}
\def\thefootnote{\fnsymbol{footnote}}

\begin{document}

\vspace{0.2cm}

\begin{center}
{\Large\bf Update on two-zero textures of the Majorana neutrino mass matrix \\ in light of recent T2K, Super-Kamiokande and NO$\nu$A results}
\end{center}

\vspace{0.2cm}

\begin{center}
{{\bf Shun Zhou $^{a,b}$} \footnote{zhoush@ihep.ac.cn}}
\\
{$^a$Institute of High Energy Physics, Chinese Academy of
Sciences, Beijing 100049, China \\
$^b$Center for High Energy Physics, Peking University, Beijing 100080, China}
\end{center}

\vspace{1.5cm}
\begin{abstract}
The latest results from atmospheric and accelerator neutrino experiments indicate that the normal neutrino mass ordering $m^{}_1 < m^{}_2 < m^{}_3$, a maximal leptonic CP-violating phase $\delta = 270^\circ$ and the second octant of neutrino mixing angle $\theta^{}_{23} > 45^\circ$ are favored. In light of new experimental results, we update previous phenomenological studies on two-zero textures of the Majorana neutrino mass matrix $M^{}_\nu$, in the flavor basis where the charged-lepton mass matrix $M^{}_l$ is diagonal. When the $1\sigma$ ranges of neutrino mixing parameters are taken into account, only four (i.e., ${\bf A}^{}_{1, 2}$ and ${\bf B}^{}_{2,4}$) among seven two-zero patterns of $M^{}_\nu$ show the aforementioned features of neutrino mass spectrum, mixing angle $\theta^{}_{23}$ and CP-violating phase $\delta$, and thus are compatible with the latest neutrino oscillation data. The correlative relations among neutrino masses and mixing parameters have been derived analytically for these four patterns, and the allowed regions of neutrino mixing angles and the CP-violating phase are also given. Possible realizations of four viable two-zero textures via non-Abelian discrete flavor symmetries are discussed.
\end{abstract}

\begin{flushleft}
\hspace{0.8cm} PACS number(s): 14.60.Pq, 11.30.Hv
\end{flushleft}

\def\thefootnote{\arabic{footnote}}
\setcounter{footnote}{0}

\newpage

\section{Introduction}

In 2011, the accelerator neutrino experiments T2K~\cite{Abe:2011sj} and MINOS~\cite{Adamson:2011qu} observed neutrino oscillations in the $\nu^{}_\mu \to \nu^{}_e$ appearance channel, indicating a nonzero $\theta^{}_{13}$ at the $2\sigma$ significance level, which is also consistent with the result from the reactor neutrino experiment Double Chooz~\cite{Abe:2011fz}. In 2012, the Daya Bay collaboration~\cite{An:2012eh} observed the disappearance of $\overline{\nu}^{}_e$ from nuclear reactors and discovered a relatively large value $\theta^{}_{13} \approx 9^\circ$ with a high statistical significance, which was further confirmed by the RENO~\cite{Ahn:2012nd} and Double Chooz~\cite{Abe:2012tg} experiments. Thanks to three leading reactor neutrino experiments, Daya Bay, Double Chooz and RENO, the value of the smallest mixing angle $\theta^{}_{13}$ has been measured more precisely than those of the other two large mixing angles $\theta^{}_{12}$ and $\theta^{}_{23}$. The remaining important issues in experimental neutrino physics are to determine whether neutrino mass ordering is normal (i.e., $m^{}_1 < m^{}_2 < m^{}_3$, denoted as NO) or inverted (i.e., $m^{}_3 < m^{}_1 < m^{}_2$, denoted as IO), to discover leptonic CP violation and measure the CP-violating phase $\delta$, and to pin down the octant of $\theta^{}_{23}$ (i.e., $\theta^{}_{23} < 45^\circ$ or $\theta^{}_{23} > 45^\circ$). Any important experimental progress will shed light on the underlying flavor structure of lepton mass matrices, and thus help us explore the origin of neutrino masses and flavor mixing.

Very recently, the T2K collaboration has published a combined analysis of the data in both $\nu^{}_\mu \to \nu^{}_\mu$ disappearance and $\nu^{}_\mu \to \nu^{}_e$ appearance channels~\cite{Abe:2015awa}. If the reactor neutrino data relevant for $\theta^{}_{13}$ are included, the likelihood maximum is reached at $\delta = 270^\circ$ and  $\sin^2 \theta^{}_{23} = 0.528$ (i.e., $\theta^{}_{23} = 46.6^\circ$) in the NO case. In addition, the latest preliminary results of $\nu^{}_\mu \to \nu^{}_e$ appearance from the NO$\nu$A experiment also favor NO at the $2\sigma$ level, and point to $\delta \approx 270^\circ$ if $\theta^{}_{23} = 45^\circ$ is fixed~\cite{Bian}. Apart from the long-baseline accelerator neutrino experiments T2K and NO$\nu$A, the atmospheric neutrino experiment Super-Kamiokande has presented the latest results on neutrino oscillation parameters~\cite{Kachulis}, showing a weak preference for NO, a nearly maximal CP-violating phase $\delta = 240^\circ$, and the second octant $\theta^{}_{23} > 45^\circ$. The preference for NO arises from an excess of up-going $\nu^{}_e$-like events, which could be interpreted by the resonant flavor conversion associated with $\theta^{}_{13}$ in the neutrino channel. The combined analysis of accelerator and reactor neutrino data has been performed in Ref.~\cite{Elevant:2015ska}, where the values of $\delta$ around $90^\circ$ are found to be disfavored at between $2\sigma$ and $3\sigma$, depending on neutrino mass ordering and the octant of $\theta^{}_{23}$. Similar conclusions are also reached in Ref.~\cite{Palazzo:2015gja}. Although the statistical significance of the above results is below $3\sigma$ for each experiment, we have the same indication from both accelerator and atmospheric neutrino oscillation experiments, for which the neutrino sources, oscillation physics and detection methods are completely different. Therefore, it is very interesting to study the implications of a combination of NO, $\delta = 270^\circ$ and $\theta^{}_{23} > 45^\circ$ for the flavor structure of the Majorana neutrino mass matrix.\footnote{In the present work, neutrinos are assumed to be Majorana particles, as they indeed are in a class of seesaw models for neutrino mass generation.} Based on the possible flavor structure, one can go further to explore the dynamical origin of neutrino masses and underlying flavor symmetries.

In this paper, we update the phenomenological studies of two-zero textures of Majorana neutrino mass matrix in Ref.~\cite{Fritzsch:2011qv}. The motivation of such an update is two-fold. First, in the flavor basis where the flavor eigenstates of charged leptons coincide with their mass eigenstates, it is straightforward to show that two-zero textures of Majorana neutrino mass matrix contain only five real parameters, which should be confronted with nine observables, including three neutrino masses, three flavor mixing angles, and three CP-violating phases. As a consequence, they are very predictive, and can readily be tested in neutrino oscillation experiments. Second, Ref.~\cite{Fritzsch:2011qv} and others~\cite{Ludl:2011vv,Meloni:2012sx} were inspired by the discovery of a nonzero $\theta^{}_{13}$, so they did not focus on the mass ordering, a maximal CP-violating phase or the octant of $\theta^{}_{23}$. In particular, only the parameter space around $\delta = 90^\circ$ or satisfiying $\theta^{}_{23} < 45^\circ$ is concentrated on in Ref.~\cite{Fritzsch:2011qv}, which is now inappropriate in light of recent T2K, Super-Kamiokande and NO$\nu$A results. Furthermore, we have derived a novel analytical formula for the correlation among neutrino mixing parameters, which is applicable to a special set of two-zero textures and turns out to be much simpler than that existing in the literature. Among seven two-zero textures, we find that only four (i.e., ${\bf A}^{}_{1,2}$ and ${\bf B}^{}_{2,4}$) remain consistent with the latest neutrino oscillation data if the $1\sigma$ ranges of mixing parameters are taken into account.

The rest of this paper is structured as follows. In Sec. 2, the important and basic formulas for two-zero textures are summarized. Sec. 3 is devoted to analytical and approximate results of theoretical predictions, and a complete numerical analysis of the allowed parameter space. We have found a novel and useful formula that has not yet been noticed in previous works. It turns out that only four out of seven patterns (i.e., ${\bf A}^{}_{1, 2}$ and ${\bf B}^{}_{2,4}$) are compatible with the latest neutrino oscillation data. The precise determination of $\delta$ and $\theta^{}_{23}$ could further single out a unique pattern of two-zero textures, or exclude all of them eventually. In Sec. 4, we present some examples to illustrate how to realize those two-zero textures through non-Abelian discrete flavor symmetries. Finally, our main conclusions are summarized in Sec. 5.

\section{Two-zero Textures}

Since neutrino masses cannot be accommodated within the standard model of elementary particles (SM), one can regard the SM as an effective theory at the electroweak scale and introduce higher-dimensional operators, which fully comply with the SM gauge symmetries. As already pointed out by Weinberg~\cite{Weinberg:1979sa}, the lowest-dimensional operator composed of lepton and Higgs doublets is $(\overline{\ell^{}_{\rm L}} \Phi) (\Phi^{\rm T} \ell^{\rm C}_{\rm L})$ of dimension five, which is unique and leads to lepton number violation by two units. After spontaneous gauge symmetry breaking, neutrinos acquire a tiny Majorana mass term from this five-dimensional operator. Such an operator can actually be realized in renormalizable models, e.g., the seesaw models, in which three singlet right-handed neutrinos~\cite{Minkowski:1977sc,Yanagida:1979ss,Gell-Mann:1979ss,Glashow:1979ss,Mohapatra:1979ia} or one triplet scalar~\cite{type2,Valle2,Cheng,Magg,Shafi,Mohapatra} or three triplet fermions~\cite{type3} are introduced and the mass scale of these new particles is extremely high.

In the seesaw framework, the mass terms relevant for lepton mass spectra and flavor mixing at low-energy scales can be written as
\begin{equation}
{\cal L}^{}_{\rm m} = - \overline{l^{}_{\rm L}} M^{}_l E^{}_{\rm R} - \frac{1}{2} \overline{\nu^{}_{\rm L}} M^{}_\nu \nu^{\rm C}_{\rm L} + {\rm h.c.} \; ,
%     (1)
\end{equation}
where $M^{}_l$ and $M^{}_\nu$ stand for the charged-lepton and Majorana neutrino mass matrix, respectively. In the flavor basis, where $M^{}_l = {\rm diag}\left\{m^{}_e, m^{}_\mu, m^{}_\tau\right\}$ is diagonal, the lepton flavor mixing matrix $V$ stems from the diagonalization of a complex and symmetric matrix $M^{}_\nu$, namely, $V^\dagger M^{}_\nu V^* = \widehat{M}^{}_\nu \equiv {\rm diag}\left\{m^{}_1, m^{}_2, m^{}_3\right\}$. Here $m^{}_\alpha$ (for $\alpha = e, \mu, \tau$) and $m^{}_i$ (for $i = 1, 2, 3$) denote respectively the charged-lepton and neutrino masses. For later convenience, we decompose the mixing matrix as $V = U\cdot P$, where $P \equiv {\rm diag}\{e^{{\rm i}\rho}, e^{{\rm i}\sigma}, 1\}$ with $\{\rho, \sigma\}$ being two Majorana CP-violating phases, and $U$ can be parametrized in terms of three mixing angles $\{\theta^{}_{12}, \theta^{}_{13}, \theta^{}_{23}\}$ and one Dirac CP-violating phase $\delta$, namely,
\begin{equation}
U = \left(\begin{matrix} c^{}_{12} c^{}_{13} & s^{}_{12} c^{}_{13}
& s^{}_{13} e^{-{\rm i} \delta} \cr -s^{}_{12} c^{}_{23} - c^{}_{12} s^{}_{13} s^{}_{23} e^{{\rm i} \delta} & + c^{}_{12} c^{}_{23} - s^{}_{12} s^{}_{13} s^{}_{23} e^{{\rm i} \delta} & c^{}_{13} s^{}_{23} \cr + s^{}_{12} s^{}_{23} - c^{}_{12} s^{}_{13} c^{}_{23} e^{{\rm i} \delta} & - c^{}_{12} s^{}_{23} - s^{}_{12} s^{}_{13} c^{}_{23} e^{{\rm i} \delta} & c^{}_{13} c^{}_{23} \cr \end{matrix}\right) \; ,
%     (2)
\end{equation}
with $s^{}_{ij} \equiv \sin \theta^{}_{ij}$ and $c^{}_{ij} \equiv \cos \theta^{}_{ij}$ for $ij = 12, 13, 23$. The three mixing angles $\{\theta^{}_{12}, \theta^{}_{13}, \theta^{}_{23}\}$, two neutrino mass-squared differences $\Delta m^2_{21} \equiv m^2_2 - m^2_1$ and $\Delta m^2_{31} \equiv m^2_3 - m^2_1$, and the Dirac CP-violating phase $\delta$ can be determined from neutrino oscillation experiments. In addition, the constraints on the absolute neutrino masses can be obtained from both beta-decay and neutrinoless double-beta decay experiments, while the latter could also provide us with some useful information on the two Majorana CP-violating phases. The latest observations of cosmic microwave background by the Planck collaboration have placed a restrictive bound on the sum of neutrino masses $\Sigma = m^{}_1 + m^{}_2 + m^{}_3 < 0.23~{\rm eV}$ at the $95\%$ confidence level~\cite{Ade:2015xua}.

In this context, any useful experimental information on the mixing matrix $V$ and neutrino masses $m^{}_i$ (for $i = 1, 2, 3$) can be implemented to reconstruct the Majorana neutrino mass matrix via
\begin{equation}
M^{}_\nu = V \left(\begin{matrix} m^{}_1 & 0 & 0 \cr 0 & m^{}_2 & 0 \cr 0 & 0 & m^{}_3 \end{matrix}\right) V^{\rm T} = U \left(\begin{matrix} \lambda^{}_1 & 0 & 0 \cr 0 & \lambda^{}_2 & 0 \cr 0 & 0 & \lambda^{}_3 \end{matrix}\right) U^{\rm T} \; ,
%     (3)
\end{equation}
where $\lambda^{}_1 \equiv m^{}_1 e^{2{\rm i}\rho}$, $\lambda^{}_2 \equiv m^{}_2 e^{2{\rm i}\sigma}$, and $\lambda^{}_3 \equiv m^{}_3$. On the other hand, any constraints on $M^{}_\nu$ arising from flavor symmetries or empirical assumptions will result in specific predictions for neutrino mass spectrum and mixing parameters, which are ready to be tested in neutrino oscillation experiments. Now we follow the second way and assume two texture zeros in $M^{}_\nu$, which were first considered by Frampton, Glashow and Marfatia in Ref.~\cite{Frampton:2002yf} and thoroughly studied by Xing in a series of papers~\cite{Xing:2002ta,Xing:2002ap,Guo:2002ei}, and by some other authors~\cite{Kaneko:2002yp,Branco:2002xf,Grimus:2011sf}. The two-zero textures have been reexamined after the experimental discovery of a nonzero $\theta^{}_{13}$ in Refs.~\cite{Fritzsch:2011qv,Ludl:2011vv,Meloni:2012sx,Frigerio:2013uva,Meloni:2014yea,Felipe:2014vka,Cebola:2015dwa}.

The predictions from two-zero textures and their experimental tests can be found in previous works, however, we summarize the important formulas in this section and establish our notations. Since $M^{}_\nu$ is symmetric, there are totally six independent matrix elements. If two of them are assumed to be zero, i.e., $(M^{}_\nu)_{a b} = (M^{}_\nu)_{c d} = 0$ for $ab \neq cd$, we obtain
\begin{eqnarray}
\frac{\lambda^{}_1}{\lambda^{}_3} U^{}_{a 1} U^{}_{b 1} + \frac{\lambda^{}_2}{\lambda^{}_3} U^{}_{a 2} U^{}_{b 2} &=& - U^{}_{a 3} U^{}_{b 3} \; , \nonumber \\
\frac{\lambda^{}_1}{\lambda^{}_3} U^{}_{c 1} U^{}_{d 1} + \frac{\lambda^{}_2}{\lambda^{}_3} U^{}_{c 2} U^{}_{d 2} &=& - U^{}_{c 3} U^{}_{d 3} \; .
%     (4)
\end{eqnarray}
From Eq.~(4), one can immediately extract the neutrino mass ratios~\cite{Xing:2002ta,Xing:2002ap,Guo:2002ei,Fritzsch:2011qv}
\begin{eqnarray}
\xi &\equiv& \frac{m^{}_1}{m^{}_3} = \left| \frac{U^{}_{a3} U^{}_{b3} U^{}_{c2} U^{}_{d2} - U^{}_{a2} U^{}_{b2} U^{}_{c3} U^{}_{d3}}{U^{}_{a2} U^{}_{b2} U^{}_{c1} U^{}_{d1} - U^{}_{a1} U^{}_{b1} U^{}_{c2} U^{}_{d2}} \right| \; , \nonumber \\
\zeta &\equiv& \frac{m^{}_2}{m^{}_3} = \left| \frac{U^{}_{a1} U^{}_{b1} U^{}_{c3} U^{}_{d3} - U^{}_{a3} U^{}_{b3} U^{}_{c1} U^{}_{d1}}{U^{}_{a2} U^{}_{b2} U^{}_{c1} U^{}_{d1} - U^{}_{a1} U^{}_{b1} U^{}_{c2} U^{}_{d2}}  \right| \; ;
\end{eqnarray}
and two Majorana CP-violating phases
\begin{eqnarray}
\rho &=& \frac{1}{2} \arg \left[ \frac{U^{}_{a3} U^{}_{b3} U^{}_{c2} U^{}_{d2} - U^{}_{a2} U^{}_{b2} U^{}_{c3} U^{}_{d3}}{U^{}_{a2} U^{}_{b2} U^{}_{c1} U^{}_{d1} - U^{}_{a1} U^{}_{b1} U^{}_{c2} U^{}_{d2}} \right] \; , \nonumber \\
\sigma &=& \frac{1}{2} \arg \left[ \frac{U^{}_{a1} U^{}_{b1} U^{}_{c3} U^{}_{d3} - U^{}_{a3} U^{}_{b3} U^{}_{c1} U^{}_{d1}}{U^{}_{a2} U^{}_{b2} U^{}_{c1} U^{}_{d1} - U^{}_{a1} U^{}_{b1} U^{}_{c2} U^{}_{d2}} \right] \; .
\end{eqnarray}

Now it is instructive to count the number of free parameters in this scenario. The physical parameters include three neutrino masses $\{m^{}_1, m^{}_2, m^{}_3\}$, three mixing angles $\{\theta^{}_{12}, \theta^{}_{13}, \theta^{}_{23}\}$ and three CP-violating phases $\{\delta, \rho, \sigma\}$. Given the four constraints or relations in Eqs. (5) and (6), we are left with five degrees of freedom, which can be fixed by two neutrino mass-squared differences $\Delta m^2_{21}$ and $\Delta m^2_{31}$, and three mixing angles $\{\theta^{}_{12}, \theta^{}_{13}, \theta^{}_{23}\}$. Therefore, the absolute neutrino mass (e.g., $m^{}_3$) and CP-violating phases $\{\delta, \rho, \sigma\}$ can be determined by the available neutrino oscillation data on $\{\Delta m^2_{21}, \Delta m^2_{31}\}$ and $\{\theta^{}_{12}, \theta^{}_{13}, \theta^{}_{23}\}$. To be explicit, neutrino masses can be found by using
\begin{equation}
m^{}_3 = \sqrt{\frac{\Delta m^2_{21}}{\zeta^2 - \xi^2}} \; , ~~~~~~~~ m^{}_2 = \zeta m^{}_3 \; , ~~~~~~~~ m^{}_1 = \xi m^{}_3 \; ,
\end{equation}
and the Dirac CP-violating phase $\delta$ can be fixed by
\begin{equation}
R^{}_\nu \equiv \frac{\Delta m^2_{21}}{|\Delta m^2_{31}|} = \frac{\zeta^2 - \xi^2}{|1 - \xi^2|} \; ,
\end{equation}
where the dependence of $\zeta$ and $\xi$ on $\delta$ is implied in Eq.~(5). The approximate analytical results of $\{\delta, \rho, \sigma\}$ can be found in Refs.~\cite{Xing:2002ta,Xing:2002ap,Guo:2002ei,Fritzsch:2011qv}. In Ref.~\cite{Fritzsch:2011qv}, the global-fit results of neutrino oscillation data at that time have been used to check the viability of two-zero textures of $M^{}_\nu$. It turns out that the following seven patterns survive current neutrino oscillation data:
\begin{eqnarray}
&& {\bf A}^{}_1: \left(\begin{matrix} {\bf 0} & {\bf 0} & a \cr
                                   {\bf 0} & b & c \cr
                                   a & c & d \cr \end{matrix}\right) \; , ~~                              {\bf A}^{}_2:
              \left(\begin{matrix} {\bf 0} & a & {\bf 0} \cr
                                   a & b & c \cr
                                   {\bf 0} & c & d \cr \end{matrix}\right) \; , ~~ {\bf B}^{}_1:
              \left(\begin{matrix} a & b & {\bf 0} \cr
                                   b & {\bf 0} & c \cr
                                   {\bf 0} & c & d \cr \end{matrix}\right) \; , ~~                               {\bf B}^{}_2:
              \left(\begin{matrix} a & {\bf 0} & b \cr
                                   {\bf 0} & c & d \cr
                                   b & d & {\bf 0} \cr \end{matrix}\right) \; , ~~
                                   \\
&& {\bf B}^{}_3: \left(\begin{matrix} a & {\bf 0} & b \cr
                                   {\bf 0} & {\bf 0} & c \cr
                                    b& c & d \cr \end{matrix}\right) \; , ~~                               {\bf B}^{}_4:
              \left(\begin{matrix} a & b & {\bf 0} \cr
                                   b & c & d \cr
                                   {\bf 0} & d & {\bf 0} \cr \end{matrix}\right) \; , ~~~~ {\bf C}: ~~~~~~~~
              \left(\begin{matrix} a & b & c \cr
                                   b & {\bf 0} & d \cr
                                   c & d & {\bf 0} \cr \end{matrix}\right) \; ,
\end{eqnarray}
where the nonzero elements are in general complex. However, it is always possible to make three nonzero matrix elements real by redefining the phases of three charged-lepton fields. In this case, we have five free model parameters (e.g., three real elements $\{a, b, c\}$ and one complex element $d$), as it should be consistent with our parameter counting.
%%%%%%%%%%%%%%%%%%%%%%%%%% Table 1 %%%%%%%%%%%%%%%%%%%%%%%%%%%%
%%%%%%%%%%%%%%%%%%%%%%%%%%%%%%%%%%%%%%%%%%%%%%%%%%%%%%%%%%%%%%%%%
\begin{table}[t]
\begin{center}
\vspace{-0.25cm} \caption{The best-fit values, together with the
1$\sigma$, 2$\sigma$ and 3$\sigma$ intervals, for three neutrino
mixing angles $\{\theta^{}_{12}, \theta^{}_{13}, \theta^{}_{23}\}$, two mass-squared differences $\{\Delta m^2_{21}, \Delta m^2_{31}~{\rm or}~\Delta m^2_{32}\}$ and the Dirac CP-violating phase $\delta$ from a global analysis of current experimental data~\cite{Gonzalez-Garcia}. Two independent global-fit analyses can be found in Refs.~\cite{Fogli,Valle}, which are in perfect agreement with the results presented here at the $3\sigma$ level.} \vspace{0.2cm}
\begin{tabular}{c|c|c|c|c}
\hline
\hline
Parameter & Best fit & 1$\sigma$ range & 2$\sigma$ range & 3$\sigma$ range \\
\hline
\multicolumn{5}{c}{Normal neutrino mass ordering
$(m^{}_1 < m^{}_2 < m^{}_3$)} \\ \hline
%------------------------------------------------------------
$\theta_{12}/^\circ$
& $33.48$ & 32.73 --- 34.26 & 31.98 --- 35.04 & 31.29 --- 35.91 \\
%------------------------------------------------------------
$\theta_{13}/^\circ$
& $8.50$ & 8.29 --- 8.70 & 8.08 --- 8.90 & 7.85 --- 9.10 \\
%------------------------------------------------------------
$\theta_{23}/^\circ$
& $42.3$  & 40.7 --- 45.3 & 39.1 --- 48.3 & 38.2 --- 53.3 \\
%------------------------------------------------------------
$\delta/^\circ$ &  $306$ & 236 --- 345 & 0 --- 24
$\oplus$ 166 --- 360 & 0 --- 360 \\
%------------------------------------------------------------
$\Delta m^2_{21}/[10^{-5}~{\rm eV}^2]$ &  $7.50$ & 7.33 --- 7.69 & 7.16 --- 7.88 & 7.02 --- 8.09 \\
%------------------------------------------------------------
$\Delta m^2_{31}/[10^{-3}~{\rm eV}^2]$ &  $+2.457$ & +2.410 --- +2.504 & +2.363 --- +2.551 & +2.317 --- +2.607 \\\hline
%%%%%%%%%%%%%%%%%%%%%%%%%%%%%%%%%%%%%%%%%%%%%%%%%%%%%%%%%%%%%
\multicolumn{5}{c}{Inverted neutrino mass ordering
$(m^{}_3 < m^{}_1 < m^{}_2$)} \\ \hline
%------------------------------------------------------------
$\theta_{12}/^\circ$
& $33.48$ & 32.73 --- 34.26 & 31.98 --- 35.04 & 31.29 --- 35.91 \\
%------------------------------------------------------------
$\theta_{13}/^\circ$
& $8.51$ & 8.30 --- 8.71 & 8.09 --- 8.91 & 7.87 --- 9.11 \\
%------------------------------------------------------------
$\theta_{23}/^\circ$
& $49.5$  & 47.3 --- 51.0 & 45.1 --- 52.5 & 38.6 --- 53.3 \\
%------------------------------------------------------------
$\delta/^\circ$ &  $254$ & 192 --- 317 & 0 --- 20
$\oplus$ 130 --- 360 & 0 --- 360 \\
%------------------------------------------------------------
$\Delta m^2_{21}/[10^{-5}~{\rm eV}^2]$ &  $7.50$ & 7.33 --- 7.69 & 7.16 --- 7.88 & 7.02 --- 8.09 \\
%------------------------------------------------------------
$\Delta m^2_{32}/[10^{-3}~{\rm eV}^2]$ &  $-2.449$ & $-2.496$ --- $-2.401$ & $-2.543$ --- $-2.355$ & $-2.590$ --- $-2.307$ \\ \hline\hline
%%%%%%%%%%%%%%%%%%%%%%%%%%%%%%%%%%%%%%%%%%%%%%%%%%%%%%%%%%%%%
\end{tabular}
\end{center}
\end{table}
%%%%%%%%%%%%%%%%%%%%%%%%%%%%%%%%%%%%%%%%%%%%%%%%%%%%%%%%%%%%%%%%%
%%%%%%%%%%%%%%%%%%%%%%%%%%%%%%%%%%%%%%%%%%%%%%%%%%%%%%%%%%%%%%%%%

Motivated by the recent results from T2K, Super-Kamiokande and NO$\nu$A, we reexamine those seven patterns by focusing on whether NO, $\delta = 270^\circ$ and $\theta^{}_{23} > 45^\circ$ are allowed simultaneously. If this is the case for a given pattern, we further completely explore the viable regions of neutrino masses and flavor mixing parameters. Unfortunately, the latest global-fit analysis of neutrino oscillation data has not yet included the preliminary results from T2K, Super-Kamiokande and NO$\nu$A, although partial joint analyses of reactor and long-baseline accelerator neutrino data have been performed in Refs.~\cite{Elevant:2015ska,Palazzo:2015gja}. For comparison, we collect the global-fit data from Ref.~\cite{Gonzalez-Garcia} in Table 1. Due to the lack of a full statistical analysis, we quote the $1\sigma$ intervals of $\Delta m^2_{21}$ and $\theta^{}_{12}$ from Table 1 in the NO case, while the $68\%$ Bayesian credible regions of $\Delta m^2_{32}$, $\theta^{}_{23}$, $\theta^{}_{13}$ and $\delta$ are taken from Ref.~\cite{Abe:2015awa}, in which a Bayesian analysis of both disappearance and appearance data from T2K and reactor neutrino data has been performed. The oscillation data used in our numerical calculations are summarized as follows
\begin{eqnarray}
32.73^\circ \leq \theta^{}_{12} \leq 34.26^\circ \; , ~~~~~ 8.61^\circ \leq \theta^{}_{13} \leq 9.56^\circ \; , ~~~~~ 44.43^\circ \leq \theta^{}_{23} \leq 49.78^\circ \; , ~~~~~ 202^\circ \leq \delta \leq 338^\circ
%     (10)
\end{eqnarray}
and
\begin{eqnarray}
&& 7.33 \times 10^{-5}~{\rm eV}^2 \leq \Delta m^2_{21} \leq 7.69 \times 10^{-5}~{\rm eV}^2 \; , \nonumber \\
&& 2.47 \times 10^{-3}~{\rm eV}^2 \leq \Delta m^2_{31} \leq 2.69 \times 10^{-3}~{\rm eV}^2 \; .
\end{eqnarray}
The best-fit values of three mixing angles ($\theta^{}_{12} = 33^\circ$, $\theta^{}_{13} = 9^\circ$, $\theta^{}_{23} = 47^\circ$), two mass-squared differences ($\Delta m^2_{21} = 7.50\times 10^{-5}~{\rm eV}^2$, $\Delta m^2_{31} = 2.58\times 10^{-3}~{\rm eV}^2$), and the CP-violating phase $\delta = 270^\circ$ will also be used for illustration.

\section{Analytical and Numerical Results}

Before proceeding with a numerical analysis, we briefly comment on the analytical results presented in Refs.~\cite{Xing:2002ta,Xing:2002ap,Guo:2002ei,Fritzsch:2011qv}, where the standard parametrization of $U$ in Eq.~(2) has been inserted into Eqs.~(5) and (6) to derive the neutrino mass ratios and CP-violating phases in terms of mixing angles. Of course, this is a correct and straightforward way to achieve an analytical understanding of any numerical results. However, for the patterns ${\bf A}^{}_1$, ${\bf A}^{}_2$, ${\bf B}^{}_3$ and ${\bf B}^{}_4$ in Eqs. (9) and (10), we can reach the analytical results in a novel and simple way. The two zeros in these four patterns are located in the same row or column of $M^{}_\nu$, namely, $(M^{}_\nu)^{}_{aa} = (M^{}_\nu)^{}_{ab} = 0$ for $a \neq b$. In this case, Eq.~(4) can be recast into
\begin{eqnarray}
\frac{\lambda^{}_1 U^{}_{a 1}}{\lambda^{}_3 U^{}_{a 3}}  U^{}_{a 1} + \frac{\lambda^{}_2 U^{}_{a 2}}{\lambda^{}_3 U^{}_{a 3}}  U^{}_{a 2} &=& - U^{}_{a 3} \; , \nonumber \\
\frac{\lambda^{}_1 U^{}_{a 1}}{\lambda^{}_3 U^{}_{a 3}}  U^{}_{b 1} + \frac{\lambda^{}_2 U^{}_{a 2}}{\lambda^{}_3 U^{}_{a 3}}  U^{}_{b 2} &=& - U^{}_{b 3} \; ;
%     (12)
\end{eqnarray}
or equivalently
\begin{equation}
\left( \begin{matrix} U^{}_{a 1} & U^{}_{a 2} \cr U^{}_{b 1} & U^{}_{b 2} \cr \end{matrix} \right) \left(\begin{matrix} \displaystyle \frac{\lambda^{}_1 U^{}_{a 1}}{\lambda^{}_3 U^{}_{a 3}} \cr ~ \cr \displaystyle \frac{\lambda^{}_2 U^{}_{a 2}}{\lambda^{}_3 U^{}_{a 3}} \cr \end{matrix}\right) = - \left(\begin{matrix} U^{}_{a 3} \cr U^{}_{b 3}\end{matrix}\right) \; .
%     (13)
\end{equation}
Then, the orthogonality conditions of the unitary matrix $U$ bring us another array of equations, which can be put in a form similar to Eq.~(14), i.e.,
\begin{equation}
\left( \begin{matrix} U^{}_{a 1} & U^{}_{a 2} \cr U^{}_{b 1} & U^{}_{b 2} \cr \end{matrix} \right) \left(\begin{matrix} \displaystyle \frac{U^*_{c 1}}{U^*_{c 3}} \cr ~ \cr \displaystyle \frac{U^*_{c 2}}{U^*_{c 3}} \cr \end{matrix}\right) = - \left(\begin{matrix} U^{}_{a 3} \cr U^{}_{b 3}\end{matrix}\right) \; ,
%     (14)
\end{equation}
with $c \neq a \neq b$. If $U^{}_{a1} U^{}_{b2} - U^{}_{a2} U^{}_{b1} \neq 0$, which is generally satisfied for the allowed values of mixing parameters, the solutions to Eq.~(14) and Eq.~(15) should be identical. By identifying these two solutions, we obtain a set of new relations
\begin{eqnarray}
\frac{\lambda^{}_1}{\lambda^{}_3} = \frac{U^*_{c 1} U^{}_{a 3}}{U^*_{c 3} U^{}_{a1}} \; , ~~~~~~~~~\hspace{0.2cm}
\frac{\lambda^{}_2}{\lambda^{}_3} = \frac{U^*_{c 2} U^{}_{a 3}}{U^*_{c 3} U^{}_{a2}} \; ,
%     (15)
\end{eqnarray}
from which one can extract two neutrino mass ratios $\{\xi, \zeta\}$ and two Majorana CP-violating phases $\{\rho, \sigma\}$ in the same manner as in Eqs.~(5) and (6). Obviously, the formulas in Eq.~(16) are much simpler than those derived in previous works~\cite{Xing:2002ta,Xing:2002ap,Guo:2002ei,Fritzsch:2011qv}.

As proved in Ref.~\cite{Fritzsch:2011qv}, the predictions from ${\bf A}^{}_1$ and ${\bf A}^{}_2$ are related by a permutation symmetry corresponding to $\theta^{}_{23} \to 90^\circ - \theta^{}_{23}$, and likewise for ${\bf B}^{}_3$ and ${\bf B}^{}_4$. Thus, we apply Eq.~(16) only to ${\bf A}^{}_1$ and ${\bf B}^{}_4$, and discuss their phenomenological implications.
\begin{itemize}
\item Pattern ${\bf A}^{}_1$ with $(M^{}_\nu)^{}_{ee} = (M^{}_\nu)^{}_{e\mu} = 0$. For this pattern, we can specify the flavor indices in Eq.~(16) as $a = e$, $b = \mu$ and $c = \tau$. Consequently, one arrives at
    \begin{eqnarray}
    \frac{\lambda^{}_1}{\lambda^{}_3} &=& \frac{U^*_{\tau 1} U^{}_{e 3}}{U^*_{\tau 3} U^{}_{e1}} = + \frac{s^{}_{13}}{c^2_{13}} \left(\frac{s^{}_{12} s^{}_{23}}{c^{}_{12} c^{}_{23}} e^{{\rm i}\delta} - s^{}_{13}\right) e^{-2{\rm i}\delta} \; , \nonumber \\
    \frac{\lambda^{}_2}{\lambda^{}_3} &=& \frac{U^*_{\tau 2} U^{}_{e 3}}{U^*_{\tau 3} U^{}_{e2}} = - \frac{s^{}_{13}}{c^2_{13}} \left(\frac{c^{}_{12} s^{}_{23}}{s^{}_{12} c^{}_{23}} e^{{\rm i}\delta} + s^{}_{13}\right) e^{-2{\rm i}\delta} \; ,
    \end{eqnarray}
    which are identical to those found in Ref.~\cite{Fritzsch:2011qv} up to different definitions of two Majorana CP-violating phases. To the leading order of $s^{}_{13}$, it has been found~\cite{Xing:2002ta}
    \begin{eqnarray}
    \xi &\approx& \tan \theta^{}_{23} \tan \theta^{}_{12} \sin \theta^{}_{13} \; , \nonumber \\
    \zeta &\approx& \tan \theta^{}_{23} \cot \theta^{}_{12} \sin \theta^{}_{13} \; .
    \end{eqnarray}
    With the help of Eqs.~(11) and (18), one can get $0.98 \leq \tan \theta^{}_{23} \leq 1.2$, $0.64 \leq \tan \theta^{}_{12} \leq 0.68$ and $0.15 \leq \sin \theta^{}_{13} \leq 0.17$ at the $1\sigma$ level, implying that $\xi < \zeta < 1$, namely, NO is favored. To the order of $s^2_{13}$, both $\xi$ and $\zeta$ receive contributions from $\cos \delta$, so we can compute $\delta$ from Eq.~(8), namely,
    \begin{equation}
    \cos \delta \approx \frac{\tan \theta^{}_{23}}{\tan 2\theta^{}_{12} \sin \theta^{}_{13}} \left( \frac{\sin 2\theta^{}_{12} \tan 2\theta^{}_{12}}{4\tan^2 \theta^{}_{23} \sin^2 \theta^{}_{13}} R^{}_\nu - 1\right) \; .
    \end{equation}
    It is now evident that $\delta = 270^\circ$ corresponds to
    \begin{equation}
    R^{}_\nu = \frac{4 \tan^2 \theta^{}_{23} \sin^2\theta^{}_{13}}{\sin 2\theta^{}_{12} \tan 2\theta^{}_{12}} \; .
    \end{equation}
    Given the best-fit values of $\Delta m^2_{21} = 7.50\times 10^{-5}~{\rm eV}^2$ and $\Delta m^2_{31} = 2.58 \times 10^{-3}~{\rm eV}^2$ below Eq.~(12), one obtains $R^{}_\nu = 0.029$. Using the best-fit values of three mixing angles, we immediately find that the value on the right-hand side of Eq.~(20) is $0.014$, which is smaller than $R^{}_\nu$ by a factor of two. Therefore, the best-fit values seem not to be consistent with the pattern ${\bf A}^{}_1$.

    As we have mentioned before, the results for ${\bf A}^{}_2$ can be derived from those for ${\bf A}^{}_1$ by a replacement $\theta^{}_{23} \to 90^\circ - \theta^{}_{23}$. In this case, it is easy to verify that ${\bf A}^{}_2$ is compatible with NO as well. For $\delta = 270^\circ$, $\tan^2 \theta^{}_{23}$ on the right-hand side of Eq.~(20) will be replaced by $\cot^2 \theta^{}_{23}$, implying a even smaller $R^{}_\nu$. Hence, we conclude that both ${\bf A}^{}_1$ and ${\bf A}^{}_2$ are consistent with NO, but they may run into problems if $\delta = 270^\circ$ is experimentally confirmed. However, if the current uncertainty on $\delta$ is taken into account, both ${\bf A}^{}_1$ and ${\bf A}^{}_2$ are not yet excluded.

\item Pattern ${\bf B}^{}_4$ with $(M^{}_\nu)^{}_{\tau \tau} = (M^{}_\nu)^{}_{e\tau} = 0$. For this pattern, setting $a = \tau$, $b = e$ and $c = \mu$ in Eq.~(16), we arrive at
    \begin{eqnarray}
    \frac{\lambda^{}_1}{\lambda^{}_3} &=& \frac{U^*_{\mu 1} U^{}_{\tau 3}}{U^*_{\mu 3} U^{}_{\tau 1}} = - \frac{c^{}_{23}}{s^{}_{23}} \frac{s^{}_{12} c^{}_{23} + c^{}_{12} s^{}_{23} s^{}_{13} e^{-{\rm i} \delta}}{s^{}_{12} s^{}_{23} - c^{}_{12} c^{}_{23} s^{}_{13} e^{+{\rm i} \delta}}\; , \nonumber \\
    \frac{\lambda^{}_2}{\lambda^{}_3} &=& \frac{U^*_{\mu 2} U^{}_{\tau 3}}{U^*_{\mu 3} U^{}_{\tau 2}} = - \frac{c^{}_{23}}{s^{}_{23}} \frac{ c^{}_{12} c^{}_{23} - s^{}_{12} s^{}_{23} s^{}_{13} e^{-{\rm i} \delta}}{c^{}_{12} s^{}_{23} + s^{}_{12} c^{}_{23} s^{}_{13} e^{+{\rm i} \delta}} \; .
    \end{eqnarray}
     At the leading order, the neutrino mass ratios are given by $\xi \approx \zeta \approx \cot^2 \theta^{}_{23}$, implying NO for $\theta^{}_{23} > 45^\circ$. As for ${\bf B}^{}_3$, the analytical results can be obtained by replacing $\theta^{}_{23}$ with $90^\circ - \theta^{}_{23}$ in Eq.~(21), which implies $\xi \approx \zeta \approx \tan^2 \theta^{}_{23}$ at the leading order. It is then straightforward to observe that only IO is allowed for $\theta^{}_{23} > 45^\circ$. Therefore, we conclude that ${\bf B}^{}_3$ is now incompatible with the observed neutrino mass spectrum. For ${\bf B}^{}_4$, to the next-to-leading order, one can derive~\cite{Fritzsch:2011qv}
     \begin{equation}
     \frac{m^{}_2}{m^{}_3} - \frac{m^{}_1}{m^{}_3} \approx - \frac{4\cot^2 \theta^{}_{23} \sin \theta^{}_{13}}{\sin 2\theta^{}_{12} \sin 2\theta^{}_{23}} \cos \delta \; ,
     \end{equation}
     from which $\cos \delta < 0$ is obtained because of $m^{}_2 > m^{}_1$. Furthermore, the Dirac CP-violating phase is determined by
     \begin{eqnarray}
     \cos \delta \approx \frac{\sin 2\theta^{}_{12} \tan^2 \theta^{}_{23}}{2\sin \theta^{}_{13} \tan 2\theta^{}_{23}} R^{}_\nu \; .
     \end{eqnarray}
     Using the best-fit values of three mixing angles and two neutrino mass-squared differences, one can get $\delta \approx 269.6^\circ$ or $90.4^\circ$, which is very close to the maximal CP-violating phase. The absolute neutrino mass scale is also fixed by neutrino mixing angles and mass-squared differences as
     \begin{equation}
     m^{}_3 = \sqrt{\frac{\Delta m^2_{31}}{1 - \cot^4 \theta^{}_{23}}} \approx 0.1~{\rm eV} \; , ~~~~~ m^{}_2 \approx m^{}_1 \approx m^{}_3 \cot^2\theta^{}_{23} \approx 0.087~{\rm eV} \; ,
     \end{equation}
     where $\theta^{}_{23} = 47^\circ$ and $\Delta m^2_{31} = 2.58\times 10^{-3}~{\rm eV}^2$ have been input. Obviously, neutrino masses are nearly degenerate, and their sum is $\Sigma \approx m^{}_3(1 + 2\cot^2 \theta^{}_{23}) \approx 0.27~{\rm eV}$, which is already in tension with the current cosmological upper bound. In summary, the allowed region of the Dirac CP-violating phase is severely constrained, namely, only $\delta \approx 90^\circ$ or $270^\circ$ is viable. If $\theta^{}_{23} \to 45^\circ$, we have $\delta \to 90^\circ$ or $270^\circ$ and $m^{}_i \to \infty$, so the upper bound on absolute neutrino masses forbids the maximal mixing angle $\theta^{}_{23} = 45^\circ$ and maximal CP-violating phase $\delta = 90^\circ$ or $270^\circ$~\cite{Grimus:2011sf}.
\end{itemize}

For other cases, Eq.~(16) is no longer valid, and the analytical formulas for $\lambda^{}_1/\lambda^{}_3$ and $\lambda^{}_2/\lambda^{}_3$ become very complicated, as one can see from Refs.~\cite{Xing:2002ta,Xing:2002ap,Fritzsch:2011qv}. To complete the list of two-zero textures that survive the latest neutrino oscillation data, we quote the analytical results for ${\bf B}^{}_2$ below
\begin{itemize}
\item Pattern ${\bf B}^{}_2$ with $(M^{}_\nu)^{}_{\tau \tau} = (M^{}_\nu)^{}_{e\mu} = 0$. For this pattern, we have
    \begin{eqnarray}
    \frac{\lambda^{}_1}{\lambda^{}_3} &=&  \frac{s^{}_{12} c^{}_{12} c^{}_{23} (2s^2_{23} s^2_{13} - c^2_{23} c^2_{13}) + s^{}_{23} s^{}_{13} (s^2_{12} c^2_{23} e^{{\rm i}\delta} + c^2_{12} s^2_{23} e^{{\rm i}\delta})}{s^{}_{12} c^{}_{12} c^{}_{23} s^2_{23} + (c^2_{12} - s^2_{12}) s^3_{23} s^{}_{13} e^{{\rm i}\delta} + s^{}_{12} c^{}_{12} c^{}_{23} s^2_{13} (1 + s^2_{23}) e^{2{\rm i}\delta}} \; , \nonumber \\
    \frac{\lambda^{}_2}{\lambda^{}_3} &=& \frac{s^{}_{12} c^{}_{12} c^{}_{23} (2s^2_{23} s^2_{13} - c^2_{23} c^2_{13}) - s^{}_{23} s^{}_{13} (c^2_{12} c^2_{23} e^{{\rm i}\delta} + s^2_{12} s^2_{23} e^{{\rm i}\delta})}{s^{}_{12} c^{}_{12} c^{}_{23} s^2_{23} + (c^2_{12} - s^2_{12}) s^3_{23} s^{}_{13} e^{{\rm i}\delta} + s^{}_{12} c^{}_{12} c^{}_{23} s^2_{13} (1 + s^2_{23}) e^{2{\rm i}\delta}}\; ,
    \end{eqnarray}
    from which one can obtain $\xi \approx \zeta \approx \cot^2 \theta^{}_{23}$ at the leading order. Hence, ${\bf B}^{}_2$ is compatible with NO for $\theta^{}_{23} > 45^\circ$, whereas $\xi \approx \zeta \approx \tan^2 \theta^{}_{23}$ is obtained for ${\bf B}^{}_1$. The latter pattern is then excluded if the hints on NO and $\theta^{}_{23} > 45^\circ$ are further confirmed by future neutrino oscillation data. At the next-to-leading order, the neutrino mass difference turns out to be
    \begin{eqnarray}
    \frac{m^{}_2}{m^{}_3} - \frac{m^{}_1}{m^{}_3} \approx  \frac{4\sin \theta^{}_{13}}{\sin 2\theta^{}_{12} \sin 2\theta^{}_{23}} \cos \delta \; ,
    \end{eqnarray}
    indicating $\cos \delta > 0$ due to $m^{}_2 > m^{}_1$. Moreover, the CP-violating phase is given by
    \begin{equation}
    \cos \delta \approx - \frac{\sin 2\theta^{}_{12}}{2\sin \theta^{}_{13} \tan 2\theta^{}_{23}} R^{}_\nu \; .
    \end{equation}
    Taking the best-fit values of neutrino mixing parameters, we get $\delta \approx 270.3^\circ$ or $89.7^\circ$. The absolute neutrino mass is the same as given in Eq.~(24), and a nearly-degenerate neutrino mass spectrum is expected. To distinguish between ${\bf B}^{}_2$ and ${\bf B}^{}_4$, we have to measure the CP-violating phase as precisely as possible, e.g., an accuracy less than $1^\circ$ is required if $\delta \approx 270^\circ$ is confirmed.
\end{itemize}

Now we perform a thorough numerical analysis of all two-zero textures of $M^{}_\nu$ by using the $1\sigma$ ranges of neutrino mixing angles, Dirac CP-violating phase and neutrino mass-squared differences in Eqs.~(11) and (12). The main strategy of our numerical analysis is as follows. (1) We randomly generate the values of $\{\theta^{}_{12}, \theta^{}_{13}, \theta^{}_{23}\}$ and $\delta$ in their $1\sigma$ ranges, and then calculate two neutrino mass ratios $\{\xi, \zeta\}$ and $R^{}_\nu$. (2) Several constraints are then chosen as criteria for whether a two-zero pattern is compatible with neutrino oscillation data. First of all, the fact of $m^{}_2 > m^{}_1$ and NO, or equivalently $\xi < \zeta < 1$, should be reproduced. Then, a pair of values for $\{\Delta m^2_{21}, \Delta m^2_{31}\}$ are randomly generated and the ratio $\Delta m^2_{21}/\Delta m^2_{31}$ is compared to $R^{}_\nu$ that has been computed from neutrino mixing angles and CP-violating phase in the first step. The difference between those two is required to be vanishingly small (e.g., $< 10^{-4}$). The final numerical results are summarized in Fig.~1, and some comments are in order:
\begin{enumerate}
\item Out of seven two-zero textures, only ${\bf A}^{}_{1,2}$ and ${\bf B}^{}_{2,4}$ are found to be consistent with the neutrino oscillation data in Eqs.~(11) and (12). In Fig.~1, the allowed regions of $\{\theta^{}_{12}, \theta^{}_{13}, \theta^{}_{23}\}$ are plotted with respect to those of $\delta$ for ${\bf A}^{}_1$ (red stars), ${\bf A}^{}_2$ (blue squares), ${\bf B}^{}_2$ (pink pluses), and ${\bf B}^{}_4$ (green crosses), where the best-fit value $\delta = 270^\circ$ and the $1\sigma$ interval $202^\circ < \delta < 338^\circ$ are represented by a dashed line and a shaded area, respectively. Different from Ref.~\cite{Fritzsch:2011qv}, the whole range of $\delta$ is now considered.

\item The patterns ${\bf A}^{}_1$ and ${\bf A}^{}_2$ are well separated in the parameter space of $\delta$. In particular, the values of $\delta$ for ${\bf A}^{}_1$ are confined to a narrow area around $\delta = 180^\circ$, i.e., the CP-conserving limit. Therefore, the sign of $\cos \delta$ (i.e., $\cos \delta < 0$ for ${\bf A}^{}_1$ and $\cos \delta > 0$ for ${\bf A}^{}_2$) is of crucial importance to discriminate between these two patterns. This is also the case for ${\bf B}^{}_2$ and ${\bf B}^{}_4$, as $\cos \delta > 0$ for ${\bf B}^{}_2$ and $\cos \delta < 0$ for ${\bf B}^{}_4$. However, in the latter case, a very precise determination of $\delta$ (e.g., less than $1^\circ$) is required.

\item As one can observe from the lower plot of Fig.~1, a joint measurement of $\theta^{}_{23}$ and $\delta$ is very useful to tell ${\bf A}^{}_2$ apart from ${\bf B}^{}_{2,4}$. For ${\bf A}^{}_2$, a nearly maximal CP-violating phase $\delta \approx 90^\circ$ or $270^\circ$ is reachable only if $\theta^{}_{23}$ takes a value close to the upper bound of its $1\sigma$ range. Another way to distinguish between ${\bf A}^{}_{1,2}$ and ${\bf B}^{}_{2,4}$ is to discover neutrinoless double-beta decays, which never take place for the former cases because of a vanishing effective neutrino mass $m^{}_{\beta \beta} = |\left(M^{}_\nu\right)^{}_{ee}| = 0$. But for ${\bf B}^{}_2$ and ${\bf B}^{}_4$, we have $m^{}_{\beta \beta} = |\left(M^{}_\nu\right)^{}_{ee}| \approx m^{}_3 \approx 0.1~{\rm eV}$, which is likely to be probed in future experiments of neutrinoless double-beta decays~\cite{Pas:2015eia,Bilenky:2014uka,Bilenky:2012qi,Rodejohann:2011mu}.
\end{enumerate}
The Dirac CP-violating phase $\delta$ serves as an important discriminator for the two-zero textures of $M^{}_\nu$. Although it seems difficult to further differentiate between ${\bf B}^{}_2$ and ${\bf B}^{}_4$ due to their similar implications at low energies, their high-energy phenomenologies in a realistic flavor model may be quite different.
%%%%%%%%%%%%%%%%%%%%%%%%%% Fig. 1 %%%%%%%%%%%%%%%%%%%%%%%%%%%%
\begin{figure}[t!]
\begin{center}
\subfigure{
\includegraphics[width=0.56\textwidth]{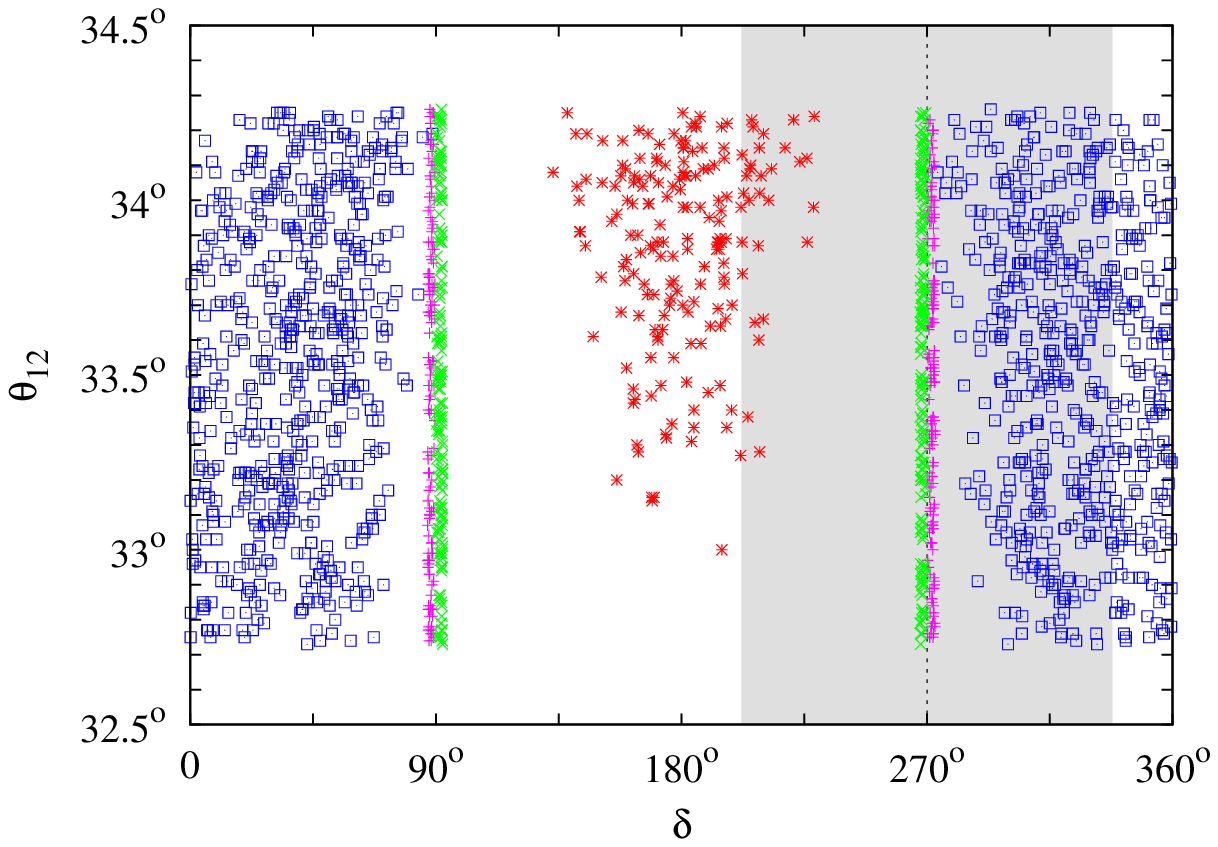} }
\subfigure{
\includegraphics[width=0.56\textwidth]{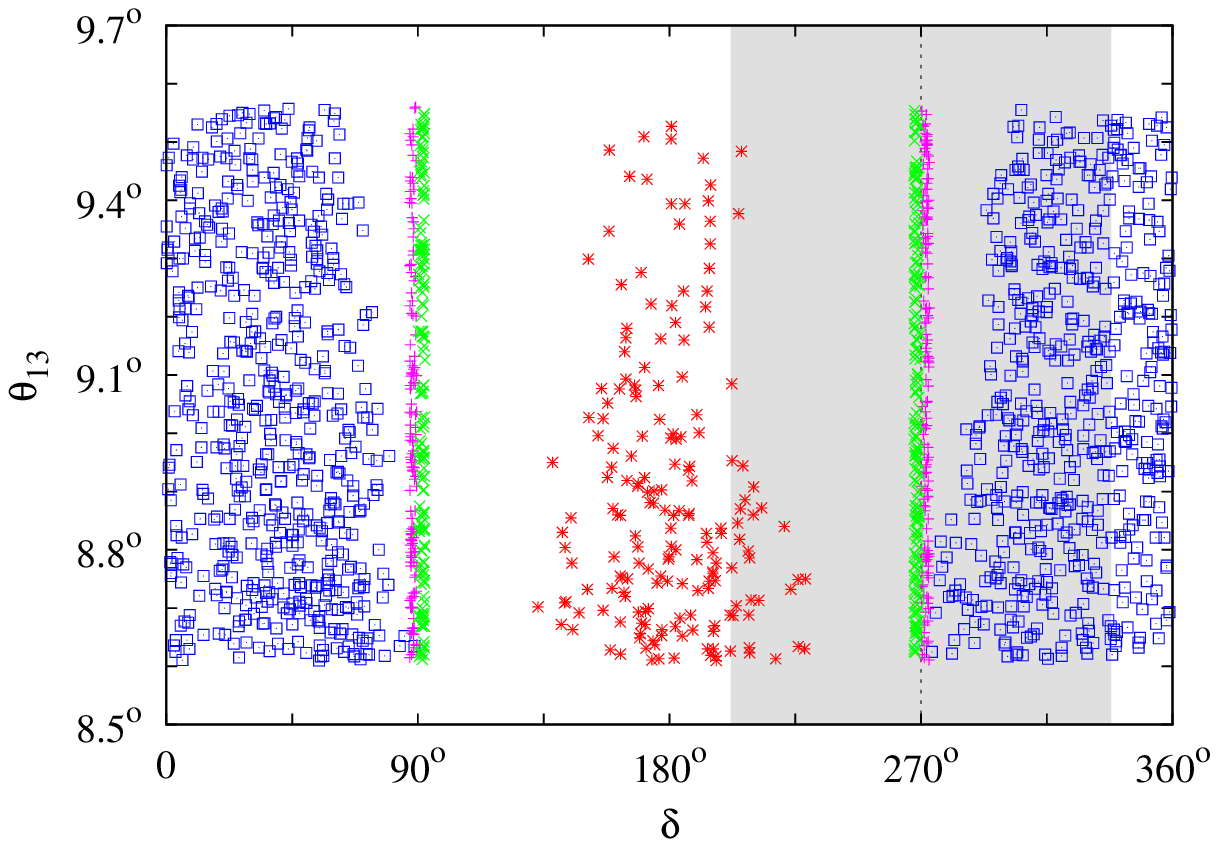} }
\subfigure{
\includegraphics[width=0.56\textwidth]{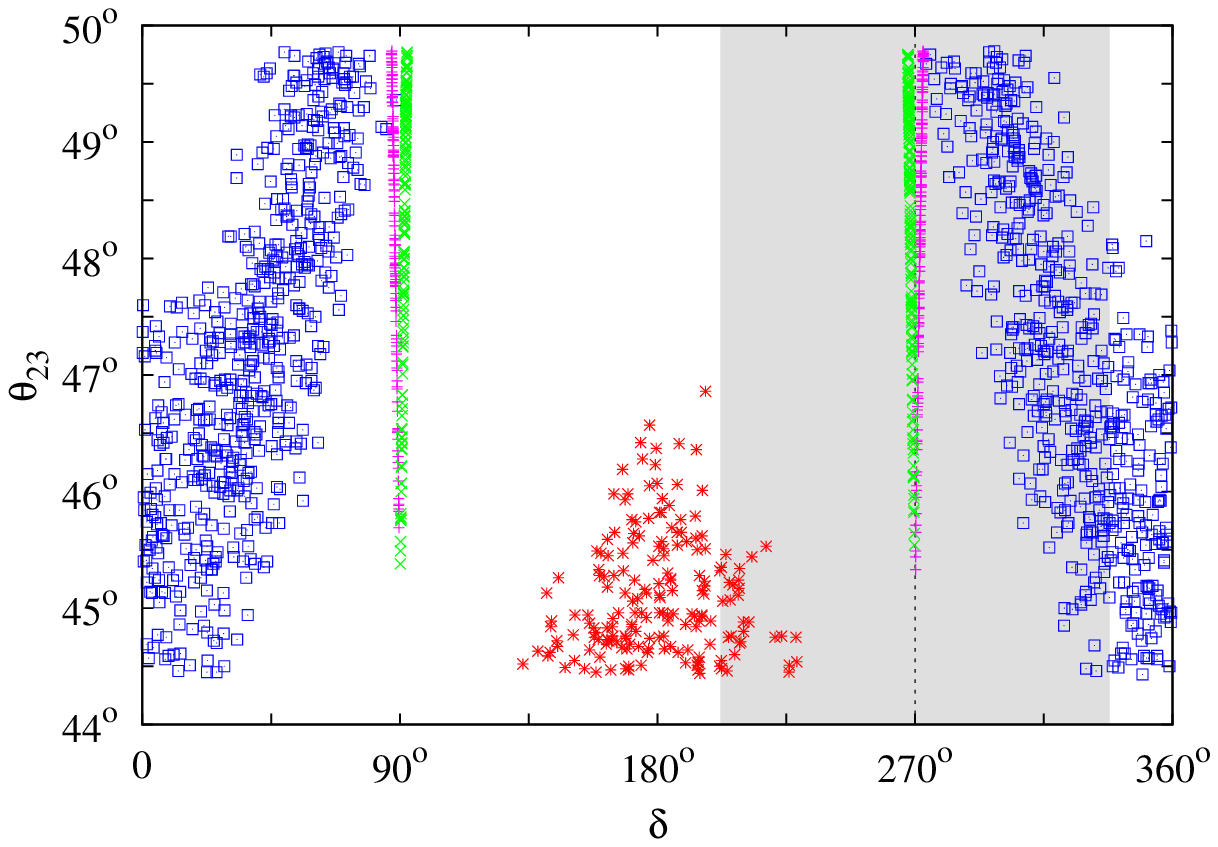} }
\caption{The allowed regions of neutrino mixing angles $\{\theta^{}_{12}, \theta^{}_{13}, \theta^{}_{23}\}$ versus the Dirac CP-violating phase $\delta$ in the scenarios of ${\bf A}^{}_1$ (red stars), ${\bf A}^{}_2$ (blue squares), ${\bf B}^{}_2$ (pink pluses), and ${\bf B}^{}_4$ (green crosses), where the shaded area corresponds to $202^\circ \leq \delta
\leq 338^\circ$ and the vertical dotted line to $\delta = 270^\circ$, and the $1\sigma$ ranges of neutrino mixing parameters in Eqs.~(11) and (12) have been used.}
\end{center}
\end{figure}
%%%%%%%%%%%%%%%%%%%%%%%%%%%%%%%%%%%%%%%%%%%%%%%%%%%%%%%%%%%%%%
\section{Symmetry Realization}

It is well known that the texture zeros in the Majorana neutrino mass matrix can always be realized by imposing Abelian flavor symmetries on the generic Lagrangian of neutrino mass models~\cite{Grimus:2004hf}. For instance, it has been demonstrated in Ref.~\cite{Fritzsch:2011qv} that cyclic symmetries $Z^{}_n$ can be used in the framework of type-II seesaw models~\cite{type2,Valle2,Cheng,Magg,Shafi,Mohapatra} to achieve all the viable two-zero textures of $M^{}_\nu$. In general, however, it is a nontrivial task to derive two-zero textures by using non-Abelian discrete flavor symmetries. A predictive neutrino mass model with an $A^{}_4$ flavor symmetry has been proposed in Ref.~\cite{Hirsch:2007kh}, where the patterns ${\bf B}^{}_1$ and ${\bf B}^{}_2$ can be obtained. In this section, we recapitulate the essential idea to produce the two-zero texture of ${\bf B}^{}_2$ in Ref.~\cite{Hirsch:2007kh}, and then propose new scenarios to realize ${\bf A}^{}_1$, ${\bf A}^{}_2$ and ${\bf B}^{}_4$.

In order to accommodate tiny neutrino masses, one can extend the SM by introducing three right-handed neutrinos $N^{}_{i{\rm R}}$ (for $i = 1, 2, 3$) and a Majorana mass term for them. To further explain lepton flavor mixing, we can implement the non-Abelian discrete flavor symmetry $A^{}_4$~\cite{Ma:2001dn,Babu:2002dz} and assign relevant fermion and scalar fields into suitable representations of the symmetry group. The assignments of fermion and scalar fields under the full symmetry ${\rm SU}(2)_{\rm L} \times {\rm U}(1)^{}_{\rm Y} \otimes A^{}_4$ are
\begin{eqnarray}
\ell^{}_{\alpha{\rm L}} \sim \left(2, -1\right) \oplus {\bf 1}, {\bf 1}^\prime, {\bf 1}^{\prime \prime} \; , E^{}_{\alpha {\rm R}} \sim \left(1, -2\right) \oplus {\bf 3} \; , N^{}_{i{\rm R}} \sim \left(1, 0\right) \oplus {\bf 3} \; , \Phi^{}_i \sim \left(2, 1\right) \oplus {\bf 3} \; , \Delta \sim \left(3, -2\right) \oplus {\bf 1}^{\prime \prime} \; ,
\end{eqnarray}
where the lepton doublets $\ell^{}_{\alpha {\rm L}}$ (for $\alpha = e, \mu, \tau$) are $A^{}_4$ singlets, three Higgs doublets $\Phi^{}_i$ form an $A^{}_4$ triplet, and one Higgs triplet $\Delta$ is an $A^{}_4$ singlet. Therefore, the gauge- and $A^{}_4$-invariant Lagrangian relevant for lepton masses and flavor mixing reads~\cite{Hirsch:2007kh}
\begin{eqnarray}
-{\cal L} &=& + a^{}_l \overline{\ell^{}_{e{\rm L}}} \left(\Phi E^{}_{\rm R}\right)^{}_{\bf 1} + b^{}_l \overline{\ell^{}_{\mu{\rm L}}} \left(\Phi E^{}_{\rm R}\right)^{}_{{\bf 1}^\prime} + c^{}_l \overline{\ell^{}_{\tau{\rm L}}} \left(\Phi E^{}_{\rm R}\right)^{}_{{\bf 1}^{\prime \prime}} \nonumber \\ && + a^{}_\nu \overline{\ell^{}_{e{\rm L}}} \left(\Phi N^{}_{\rm R}\right)^{}_{\bf 1} + b^{}_\nu \overline{\ell^{}_{\mu{\rm L}}} \left(\Phi N^{}_{\rm R}\right)^{}_{{\bf 1}^\prime} + c^{}_\nu \overline{\ell^{}_{\tau{\rm L}}} \left(\Phi N^{}_{\rm R}\right)^{}_{{\bf 1}^{\prime \prime}} \nonumber \\ && + \frac{1}{2} M \left(\overline{N^{\rm C}_{\rm R}} N^{}_{\rm R}\right)^{}_{\bf 1} + \frac{1}{2} a^{}_\Delta \left( \overline{\ell^{}_{e{\rm L}}} \Delta \ell^{\rm C}_{\tau{\rm L}}+ \overline{\ell^{}_{\tau{\rm L}}} \Delta \ell^{\rm C}_{e{\rm L}} \right) + \frac{1}{2} b^{}_\Delta \overline{\ell^{}_{\mu{\rm L}}} \Delta \ell^{\rm C}_{\mu{\rm L}} + {\rm h.c.} \; .
\end{eqnarray}
After the scalar fields acquire their vacuum expectation values $\langle \Phi^{}_i \rangle = v$ and $\langle \Delta \rangle = u$, the charged-lepton mass matrix $M^{}_l$ and Dirac neutrino mass matrix $M^{}_{\rm D}$ can be obtained from the first and second lines in Eq.~(29), respectively. The effective neutrino mass matrix $M^{}_\nu$ receives one contribution from the type-I seesaw formula $-M^{}_{\rm D} M^{\rm T}_{\rm D}/M$, and the other from the triplet Higgs. In total, we have
\begin{equation}
M^{}_\nu =  u \left(\begin{matrix}
                                   0 & 0 & a^{}_\Delta \cr
                                   0 & b^{}_\Delta & 0 \cr
                                   a^{}_\Delta & 0 & 0 \cr \end{matrix}\right) - \frac{v^2}{M} \left(\begin{matrix}
                                   a^2_\nu & 0 & 0 \cr                                  0 & 0 & b^{}_\nu c^{}_\nu \cr
                                   0 & b^{}_\nu c^{}_\nu & 0 \cr \end{matrix}\right) \; ,
\end{equation}
which can be identified as ${\bf B}^{}_2$ in Eq.~(9) with four nonzero matrix elements $a = -(a^{}_\nu v)^2/M$, $b = a^{}_\Delta u$, $c = b^{}_\Delta u$, and $d = -b^{}_\nu c^{}_\nu v^2/M$. As shown in Ref.~\cite{Hirsch:2007kh}, if the Higgs triplet is assigned as ${\bf 1}^\prime$, the pattern ${\bf B}^{}_1$ can be produced. In this model, however, it is impossible to realize any other viable two-zero textures.

To illustrate how to produce the other two-zero textures, we consider the minimal type-(I+II) seesaw model~\cite{Gu:2006wj}, where only one right-handed neutrino $N^{}_{\rm R}$ and one Higgs triplet $\Delta$ have been introduced to account for both tiny neutrino masses and baryon number asymmetry in our Universe. Now this minimal scenario is further extended, and the particle content and assignments under $A^{}_4$ symmetry are
\begin{eqnarray}
\ell^{}_{e{\rm L}} \sim {\bf 1}^{\prime \prime} \; , ~~~ \ell^{}_{\mu{\rm L}} \sim {\bf 1} \; , ~~~ \ell^{}_{\tau{\rm L}} \sim {\bf 1}^\prime \; , E^{}_{\alpha {\rm R}} \sim {\bf 3} \; , ~~~ N^{}_{\rm R} \sim {\bf 1} \; , ~~~ \Phi^{}_i \sim {\bf 3} \; , ~~~ \varphi \sim {\bf 1} \; , ~~~ \phi \sim {\bf 1}^\prime \; , ~~~ \Delta \sim {\bf 1} \; ,
\end{eqnarray}
where two Higgs doublets $\varphi$ and $\phi$ have been added as $A^{}_4$ singlets. Except for a proper rearrangement of $\overline{\ell^{}_{e{\rm L}}} \left(\Phi E^{}_{\rm R}\right)^{}_{{\bf 1}^{\prime \prime}}$, $\overline{\ell^{}_{\mu{\rm L}}}\left(\Phi E^{}_{\rm R}\right)^{}_{\bf 1}$  and $\overline{\ell^{}_{\tau{\rm L}}}\left(\Phi E^{}_{\rm R}\right)^{}_{{\bf 1}^\prime}$, the charged-lepton sector remains unchanged and the mass matrix $M^{}_l$ can be diagonalized just by rotating the right-handed charged-lepton fields in the flavor space, which has no impact on the lepton flavor mixing. The gauge- and $A^{}_4$-invariant Lagrangian relevant for neutrino masses becomes
\begin{eqnarray}
- {\cal L} = y^{}_\varphi \overline{\ell^{}_{\mu {\rm L}}} \varphi N^{}_{\rm R} + y^{}_\phi \overline{\ell^{}_{\tau {\rm L}}} \phi N^{}_{\rm R} + \frac{1}{2} M \overline{N^{\rm C}_{\rm R}} N^{}_{\rm R} + \frac{1}{2} a^{}_\Delta \left( \overline{\ell^{}_{e{\rm L}}} \Delta \ell^{\rm C}_{\tau{\rm L}}+ \overline{\ell^{}_{\tau{\rm L}}} \Delta \ell^{\rm C}_{e{\rm L}} \right) + \frac{1}{2} b^{}_\Delta \overline{\ell^{}_{\mu{\rm L}}} \Delta \ell^{\rm C}_{\mu{\rm L}} + {\rm h.c.} \; .
\end{eqnarray}
After the spontaneous symmetry breaking, the Dirac neutrino mass matrix turns out to be a $3\times 1$ column vector and is given by $M^{}_{\rm D} = \left(0, a^{}_{\rm D}, b^{}_{\rm D} \right)^{\rm T}$, where $a^{}_{\rm D} = y^{}_\varphi v^{}_\varphi$ and $b^{}_{\rm D} = y^{}_\phi v^{}_\phi$ with $v^{}_\varphi = \langle \varphi \rangle$ and $v^{}_\phi = \langle \phi \rangle$. The Majorana neutrino mass matrix is
\begin{equation}
M^{}_\nu = u \left(\begin{matrix}
                              0 & 0 & a^{}_\Delta \cr
                              0 & b^{}_\Delta & 0 \cr
                              a^{}_\Delta & 0 & 0 \cr
                              \end{matrix}\right) - \frac{1}{M} \left(\begin{matrix} 0 & 0 & 0 \cr
                              0 & a^2_{\rm D} & a^{}_{\rm D} b^{}_{\rm D}\cr
                              0 & a^{}_{\rm D} b^{}_{\rm D} & b^2_{\rm D} \cr \end{matrix}\right) \; ,
\end{equation}
which is exactly the pattern ${\bf A}^{}_1$ in Eq.~(9) with $a = a^{}_\Delta u$, $b = b^{}_\Delta u - a^2_{\rm D}/M $, $c = -a^{}_{\rm D}b^{}_{\rm D}/M$ and $d = -b^2_{\rm D}/M$. In a similar way, one can verify that the pattern ${\bf A}^{}_2$ can be achieved by assigning $\Delta$ as ${\bf 1}^{\prime \prime}$ under $A^{}_4$. Finally, to obtain ${\bf B}^{}_4$, we need to make a modification of Eq.~(31), i.e.,
\begin{eqnarray}
\ell^{}_{e{\rm L}} \sim {\bf 1} \; , ~~~ \ell^{}_{\mu{\rm L}} \sim {\bf 1}^\prime \; , ~~~ \ell^{}_{\tau{\rm L}} \sim {\bf 1}^{\prime \prime} \; , ~~~ E^{}_{\alpha {\rm R}} \sim {\bf 3} \; , ~~~ N^{}_{\rm R} \sim {\bf 1} \; , ~~~ \Phi^{}_i \sim {\bf 3} \; , ~~~ \varphi \sim {\bf 1} \; , ~~~ \phi \sim {\bf 1}^\prime \; , ~~~ \Delta \sim {\bf 1} \; ,
\end{eqnarray}
leading to a Dirac neutrino mass matrix $M^{}_{\rm D} = \left(a^{}_{\rm D}, b^{}_{\rm D}, 0\right)^{\rm T}$ and thus the Majorana neutrino mass matrix
\begin{eqnarray}
M^{}_\nu = u \left(\begin{matrix}
                              b^{}_\Delta & 0 & 0 \cr
                              0 & 0 & a^{}_\Delta \cr
                              0 & a^{}_\Delta & 0 \cr
                              \end{matrix}\right) - \frac{1}{M} \left(\begin{matrix}                               a^2_{\rm D} & a^{}_{\rm D} b^{}_{\rm D} & 0\cr
                              a^{}_{\rm D} b^{}_{\rm D} & b^2_{\rm D} & 0\cr
                              0 & 0 & 0 \end{matrix}\right) \; .
\end{eqnarray}
The phenomenological implications of these patterns for neutrino oscillation experiments have been explored in previous sections. As we have seen, it is extremely difficult to distinguish between ${\bf B}^{}_2$ and ${\bf B}^{}_4$ via neutrino oscillation experiments if the future oscillation data confirms a nearly-maximal CP-violating phase $\delta \approx 270^\circ$. For instance, the accuracy of the determination of CP-violating phase is required to be below $1^\circ$ for this purpose. In a specific model with flavor symmetries, however, new particles or interactions definitely cause additional observable effects that serve as a discriminator.

We have not tried to write down the general scalar potential that is invariant under both gauge and flavor symmetries, and investigate whether the desired vacuum expectation values can be reached. This issue can be addressed in a similar way as in a number of works on applying the $A^{}_4$ flavor symmetry to lepton flavor mixing~\cite{Altarelli:2010gt,Ishimori:2010au,King:2013eh}. Moreover, it is also interesting to examine if a successful leptogenesis works in our scenarios to explain the observed baryon number asymmetry in the Universe~\cite{Fukugita:1986hr,Davidson:2008bu}. These important questions deserve further dedicated studies, which are beyond the scope of our paper and left for future work.

\section{Summary}

In view of the latest results from atmospheric (Super-Kamiokande) and accelerator (T2K and NO$\nu$A) neutrino oscillation experiments, which hint at the normal neutrino mass ordering $m^{}_1 < m^{}_2 < m^{}_3$, the maximal CP-violating phase $\delta = 270^\circ$ and the second octant $\theta^{}_{23} > 45^\circ$, we reexamine the two-zero textures of the Majorana neutrino mass matrix. By using the $1\sigma$ ranges of neutrino oscillation parameters, we have found that only four patterns with two texture zeros (i.e., ${\bf A}^{}_{1, 2}$ and ${\bf B}^{}_{2, 4}$) pass all current experimental constraints, and they can be further tested when the precision measurements of neutrino oscillation parameters are available.

For the patterns ${\bf A}^{}_{1,2}$ and ${\bf B}^{}_{3,4}$, we have derived concise and useful formulas for neutrino mass ratios and Majorana CP-violating phases in Eq.~(16), which have not yet been noticed in the literature. The simplicity of these formulas stems from the usage of unitarity conditions, and provides an excellent explanation for why the expressions in these four cases are much simpler than those in the cases of ${\bf B}^{}_{1, 2}$ and ${\bf C}$. A detailed numerical analysis has also been carried out to check which two-zero texture is compatible with the latest experimental data. Our numerical results show that ${\bf A}^{}_1$, ${\bf A}^{}_2$, ${\bf B}^{}_2$ and ${\bf B}^{}_4$ survive the data at the $1\sigma$ level, and precise determination of $\theta^{}_{23}$ and $\delta$ is of crucial importance to discriminate one pattern from another. Some distinct features should be mentioned: (1) The sign of $\cos \delta$ is positive for ${\bf A}^{}_2$ and ${\bf B}^{}_2$, while negative for ${\bf A}^{}_1$ and ${\bf B}^{}_4$; (2) Only nearly-maximal CP-violating phase $\delta \approx 90^\circ$ or $270^\circ$ is allowed for ${\bf B}^{}_2$ and ${\bf B}^{}_4$, and the deviations are less than a few degrees; (3) The neutrino masses are nearly degenerate in the cases of ${\bf B}^{}_2$ and ${\bf B}^{}_4$, and substantial decay rates for neutrinoless double-beta decays are expected, while these decays do not occur in the cases of ${\bf A}^{}_1$ and ${\bf A}^{}_2$. Hence, in addition to oscillation experiments, the constraints on absolute neutrino masses and the observations of neutrinoless double-beta decays also help verify or disprove two-zero textures. As for model building, we have shown that the non-Abelian discrete flavor symmetry $A^{}_4$ can be used to realize these four patterns in the type-(I+II) seesaw models.

The ongoing and forthcoming neutrino oscillation experiments will unambiguously determine neutrino mass ordering and precisely measure the flavor mixing angles and the Dirac CP-violating phase. The tritium beta decays, neutrinoless double-beta decays and cosmological observations will probe the absolute neutrino mass scale and hopefully constrain the Majorana CP-violating phases. Any important experimental progress will give us new insights into the structure of lepton mass matrices, which can further help us explore the origin of neutrino masses and the true dynamics of flavor mixing and CP violation.

\section*{Acknowledgements}

The author would like to thank Y.F. Li for informing him of recent Super-Kamiokande atmospheric and NO$\nu$A results and for helpful discussions. This work was supported in part by the Innovation Program of the Institute of High Energy Physics under Grant No. Y4515570U1, by the
National Youth Thousand Talents Program, and by the CAS Center for
Excellence in Particle Physics (CCEPP).

\end{document}